# Autonomous life-like behavior emerging in active and flexible microstructures

**Authors:** Mengshi Wei[1], Daniela J. Kraft[1, *]

[1]Huygens-Kamerlingh Onnes Laboratory, Leiden University, P.O. Box 9504, 2300 RA Leiden, The Netherlands.

*Corresponding author. Email: kraft@physics.leidenuniv.nl

**Abstract:** Many organisms leverage an interplay between shape and activity to generate motion and adapt to their environment. Embedding such feedback into synthetic microrobots could eliminate the need for sensors, software, and actuators, yet current realizations are either active but rigid, or flexible but passive. Here, we introduce micrometer-scale structures that integrate both activity and flexibility by 3D microprinting concatenated units and actuating them with an AC electric field. This minimal yet versatile design gives rise to a rich array of life-like modes of motion — including railway and undulatory locomotion, rotation, and beating — as well as emergent sense-response abilities, which enable autonomous reorientation, navigation, and collision avoidance. Our approach offers a versatile platform for designing biomimetic model systems and autonomously operating microrobots with embodied intelligence.

Living organisms — from bacteria and cells to animals and humans — achieve autonomous motion by actively changing their shape. Their flexible bodies not only enable this motion but also allow for remarkable adaptability of their mode of motion across diverse environments. For instance, microscopic worms can seamlessly transition between crawling on solid substrates, swimming in liquids, and burrowing through soil (*1, 2*).

Modern robots have sought to emulate this versatility by combining activity with mechanical flexibility. Their abilities are made possible by sophisticated arrays of sensors, controllers, and actuators yet they still struggle to replicate the seamless adaption and function of their biological counterparts (*3, 4*). As robotic devices continue to shrink, life-like abilities become increasingly challenging to integrate. Most microscale designs therefore rely on external control, limiting their autonomy despite impressive technological advances (*5–9*).

Recent approaches in macroscopic soft robotics offer a promising alternative. Rather than requiring a precise temporal and spatial control of the robot, they utilize the dynamic behavior emerging in much simpler active and flexible structures. This emergent behavior enabled robust, efficient, and adaptive locomotion and shape transformations, while eliminating the need for programmed coordination and external control (*10–16*).

At the micrometer-scale, however, current structures are either active but rigid or flexible but passive. While some micrometer-sized systems have provided a glimpse of the potential for leveraging emergent dynamics at this length scale (*17–20*), they consist of disconnected units and do not exhibit the autonomous adaptive locomotion capabilities characteristic of living systems.

Here, we introduce a minimal design strategy that enables life-like motion and abilities in microscale systems. By combining activity with structural flexibility, we demonstrate that simple chain-like microstructures with feedback between their shape and motion can autonomously switch between different modes of motion and respond adaptively and efficiently to their environment – all without the need for pre-programming or external control.

**Design and fabrication of concatenated active microstructures**

To realize flexible microscopic structures, we make use of concatenation, a principle exploited in necklaces, medieval armor, and polycatenated architected materials (*21*). We materialize these concatenated structures at the micrometer scale by two-photon polymerization-based 3D microprinting. Each unit of the chain takes the form of a half cylinder with a handle attached at the front and two half disks connected by a beam at the rear end (Fig. 1, A, B and Fig. S1), which together form a hinge between adjacent units. Post-processing by UV curing and silica coating ensures stability of the sub-micrometer features and prevents sticking (Materials and Methods). Our design allows for excellent hinging between two concatenated units as reflected in their relative positions and angle distribution as measured from their thermal fluctuations (Fig. 1B). Using this principle, we can print chains made from concatenated units, with even delicate features such as the arms and hinges accurately replicated (Fig. 1C). Once released into water, their shape is subject to thermal fluctuations due to their microscopic size and flexible nature (Fig. 1C and Movie S1, Materials and Methods).

To propel the particles, we make use of an overlooked mechanism: self-dielectrophoresis (sDEP) in homogenous AC fields driven purely by the anisotropic shape of the polymer-based units (Materials and Methods) (*22–25*). When placed in an AC electric field at kHz frequencies, each anisotropic unit aligns itself along the field direction and spontaneously self-propels parallel to the substrate with its curved side leading (Fig. 1, D and E, Movie S2, Fig. S2-S4, Materials and Methods). The propulsion speed is proportional to the applied voltage $V_E{}^2$ and the propulsion and alignment persist even when the chain is lifted away from the substrate allowing 3D motion (Fig.S2 and S3). The arrangement of the units in the chain is designed such that the propulsion direction of each unit aligns with the contour of the chain, making our chains an experimental realization of tangentially driven active polymer models (*26, 27*) and models for worms (*28, 29*). The linear but flexible arrangement of active units provides high sensitivity of the motion to the shape of the chain, thereby creating a feedback mechanism. Once activated, the chain moves forward while maintaining the ability to change shape and direction (Fig. 1F, Movie S2). Dipolar repulsion between the units induces an effective bending rigidity, thus providing not only the sought-after combination of activity and flexibility but also elasticity (Fig. S5). Despite being fully synthetic, their unhindered locomotion readily evokes associations with living creatures such as the foraging and crawling motion of worms and snakes, and biological molecules such as actin filaments.

**Autonomous switching between different motion patterns**

We test whether our active and flexible structures possess a feedback between their shape and their motion as well as emergent dynamics by investigating their behavior in various conditions. type of motion, i.e., free motion on a flat substrate, displays hallmarks of a coupling between shape and motion. Upon activation, the chain straightens due to dipolar repulsions between its constituent units and undergoes 'railway motion' (*30*), i.e. each segment follows the one ahead similar to a train but without the need of a track (Fig. 2A-C and Movie S3). All units closely approach the position of the leading unit some time $\tau$ later as signified by the minimum in their relative mean squared displacement ($\Delta r^2_{1,N}$) with respect to the leading unit (Fig. 2D, Materials and Methods). Changes in the direction of the first unit thus determine the shape and locomotion of the chains.

The feedback between shape and motion furthermore enables the emergence of different, adaptive modes of motion. When the motion of the first unit is hindered by clamping (Materials and methods), the chain autonomously switches from railway-motion to spontaneous self-oscillations, reminiscent of flagellar beating (Fig. 2 E and F, Movie S4, Fig. S5). This life-like oscillatory beating relies on a combination of active forces, elastic buckling of the chain and geometric constraints: the active forces exerted by the moving units deform the chain up to the geometrically possible maximum, whose deformation in turn reorients the units thereby changing their direction of motion. This feedback between shape and motion leads to periodic oscillations and figure-of-eight patterns (Fig. 2G) (*11, 18, 26*). The beating frequency of the clamped chain $f$ can be externally tuned through the applied voltage and scales with the speed $U$ of a single unit as $f \propto U$ (Fig. 2H and 1E, Materials and Methods, Fig. S2).

To get more quantitative insights in the self-oscillations, we measure the active force through the mean polarization $\Theta = \frac{1}{N}\sum_1^N \theta_i$, where $\theta_i$ is the instantaneous orientation of the active force of unit $i \in [1, N]$ relative to the clamping direction, and the elastic buckling of the chain through the mean curvature, $\Omega = \sum_1^{N-1} \theta_{i+1} - \theta_i = \theta_N - \theta_1$. Similar to oscillating elasto-active beams at the cm length scale (*11*), we find that these parameters, which both oscillate with the beating frequency, prescribe a limit cycle, implying that the injected energy is now dissipated through the oscillation instead of through the self-propulsion (Fig. 2I, Fig. S5). The area of the limit cycle increases along the chain until it saturates at around the 8th unit, suggesting that dissipation through bending decreases along the chain, in line with the increasingly straighter shape.

The AC field activation provides further control over both the magnitude and direction of the activity (Fig. S2). By increasing the AC field frequency $f_E$ from the kHz to the MHz regime, the propulsion direction reverses such that the flat side of the half-cylinder is now leading (Fig. 2J, Fig. S2 and Movie S5). The feedback between shape and motion, however, persists, such that the active motion is now straightening the chain away from any clamping point thereby suppressing oscillations (Fig. 2K). The direction reversal can be utilized to externally control the compression and extension of clamped chains, similar to robotic arms and microorganisms like *Lacrymaria olor* (Fig. 2J-L).

Locomotion that is even more reminiscent of living microorganisms appears when the chain is freely suspended in solution but has a load attached to the front (Materials and Methods, Fig. S1) - a design that is similar to flagellated microorganisms and sperm cells. In this state, which is intermediate between free movement and clamping, the chain undergoes undulatory motion, i.e., periodic beating while moving forward (Fig. 2M and N, Movie S6), in line with predictions for tangentially driven filaments (*31*). Like clamping, the hindrance imposed by the load causes buckling of the chain, self-oscillations, and the appearance of a limit cycle (Fig. 2O). Since energy is dissipated through both propulsion and oscillations, the limit cycle is smaller. Stable oscillations, however, only occur when the load is attached symmetrically to the leading unit. Asymmetric attachment introduces circular motion, with intermittent periods of tumbling where the chain changes its shape and reverses its rotation direction (Fig. 2 P,Q,R, Movie S6).

By a simple combination of flexibility and activity, we integrated a feedback between the shape of the chain and its direction of motion, thereby imbuing it with powerful new abilities: a rich variety of modes of motion and autonomous switching between them.

**Sensing and smart adaptation to complex environments**

The embodied feedback mechanism between their conformation and their direction of motion suggests that our active and flexible microstructures should be able to sense and respond to their environment. We start testing for such a sense-respond ability by studying their motion in the presence of walls. When encountering a wall, their motion is hindered and hence leads to a buckling of the chain (Fig. 3A), similar to when becoming clamped or pushing a load. However, with the freedom to reorients all units, this buckling realigns the orientation of the leading unit, thereby changing the direction motion of the chain as a whole away from the wall (Fig. 3 A and B, Movie S7). The feedback from the buckled shape thus leads to efficient and autonomous reorientation.

We next placed them in an ordered array of immobile obstacles with spacing smaller than the length of the chain, such that they experience confinement (Materials and Methods. Fig. S1). Instead of getting trapped or hindered by the obstacles (*32–34*), our chains use the deformability of their body for easy exploration and making turns (Fig. 3 C and D, and Movie S8). Even more so, the chains sense the square pattern through the dipolar and steric repulsion from the obstacles and adapt their motion in four distinct ways: besides straight motion along the major directions, or lanes, of the array, we observe s-shapes that allow switching between lanes, right and left turns made possible by 90° bending, and U-turns which require bending by 180° (Fig. 3E, Fig. S6).

Their pliable body and sense-respond ability is not only useful to explore confined environments but also allows avoiding collisions. When two chains encounter each other in free space or narrow pathways, they sense each other's presence and deform their flexible bodies in response, allowing them to quickly bypass each other (Fig. 1F and 3F, Movie S2 and Movie S9).

Finally, we tested whether their ability to move persists even when there is little space and no clear path available, similar to worms that move through soil. To do so, we placed them in an environment densely occupied by mobile, 7 $\mu$m diameter spheres. Powered by their activity, the chains push the obstacles out of their way and are able to move in a burrowing fashion through such dense environments (Fig. 3G, Movie S10).

## Conclusions

We have developed flexible and active structures of micrometer sized concatenated units, that embody a feedback between their motion direction and shape. We have demonstrated that this feedback imbues them with different types of motion patterns and the ability to sense and adapt to their environment, leading to a life-like appearance and behavior. Being free of any tethers or other required guidance, they move and adapt fully autonomously. To showcase the functionality encompassed by a combination of flexibility and activity, we exposed the chains to obstacles as well as dense and crowded environments and found quick reorientation, easy navigation, and collision avoidance.

Our versatile and simple fabrication and activation strategy opens a virtually limitless design space to create passive and active flexible structures on the microscale with a broad array of sizes, geometries, topologies, hinging range, as well as density, location, orientation, and propulsion force of the active elements. We anticipate that this new class of active concatenated microstructures allows integration of novel functional dynamic behavior and will provide unprecedented insights into the fundamental principles governing living systems. Their emergent dynamics together with their autonomy will open new avenues in the design of biomimetic systems, biomedical applications, and intelligent microrobots with embodied intelligence, that sense and respond to their environment, autonomously take decisions, and store and process information.

**Acknowledgments:** We thank Rachel Doherty for support with 3D microprinting and Martin van Hecke for useful discussions.

**Funding:** Dutch Research Council VIDI grant 193.069 (DJK)

**Author contributions:** M.W. and D.J.K. conceived the research idea and designed the experiments; M.W. designed and fabricated the devices and conducted the experiments; M.W. and D.J.K analyzed the data, discussed the results, and wrote the paper.

**Competing interests:** The authors declare no competing interests.

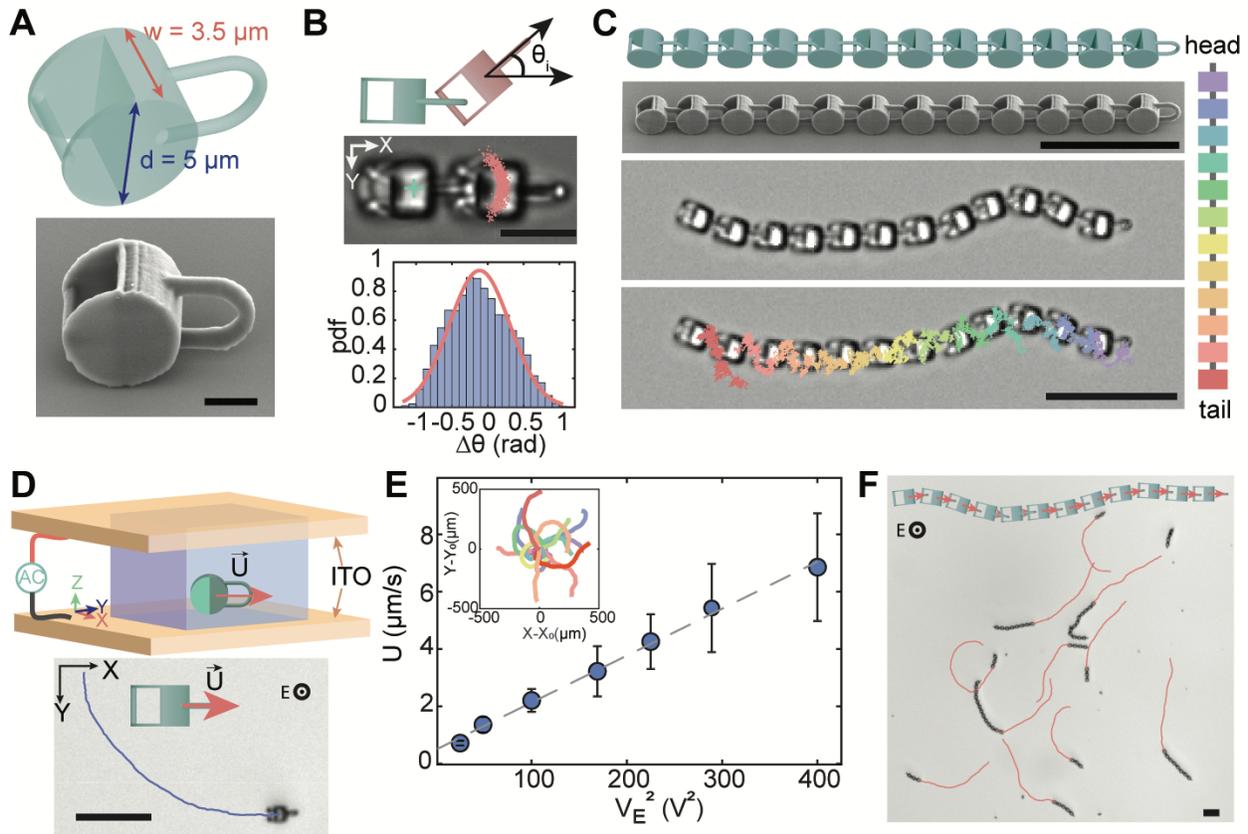

**Fig. 1. Active chains of concatenated micrometer-sized units.** (**A**) Design and SEM micrograph of a 3D microprinted anisotropic unit. Scalebar is 2 μm. (**B**) Concatenated units can move within the geometrically allowed range. Schematic with unit orientation $\theta_i$ (top) and bright field microscopy image (middle) of two units with overlaid the center of mass motions with respect to the left unit, Scalebar 5 μm; bottom: probability density function (pdf) of the relative angle $\Delta\theta = \theta_{i+1} - \theta_i$. (**C**) Realization of a flexible chain of concatenated units. From top to bottom: design, SEM image of 3D printed chain, brightfield microscopy image after release in solution and overlay with the thermally induced trajectories of each unit demonstrating the flexibility. (**D**) When applying an alternating electric field (20 V, 17 kHz), the anisotropic units move due to their anisotropic shape as visible in the brightfield snapshot with overlaid 5 s trajectory. (**E**) The velocity $U$ can be tuned by the voltage $V_E$ of the applied field. Inset: 20 s trajectories of different units centered at the origin, $f_E$ = 17 kHz. (**F**) Trajectories (10 s) of chains with different lengths show directed motion and flexibility (20 V, 10 kHz). Scalebar in C, D, F is 20 μm.

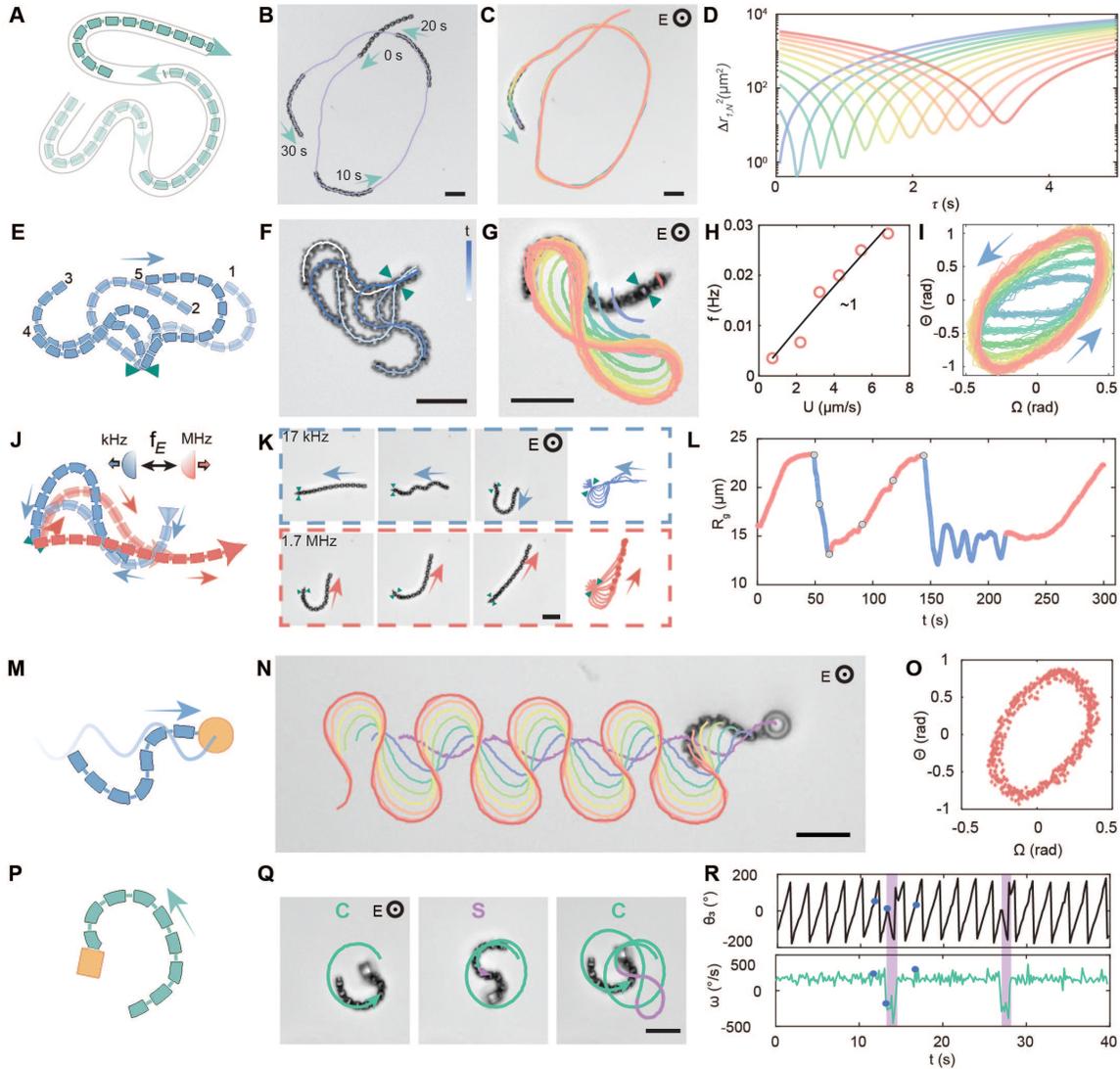

**Fig. 2. Autonomous switching between different emergent life-like modes of motion.** (**A**) - (**C**) Free 'railway' motion: (**A**) Schematic, (**B**) Overlaid microscopy snapshots with trajectory of the leading unit (purple) and (**C**) all units showing 'railway' motion, confirmed by (**D**) the relative mean squared displacement. (**E-I**) Head-clamping induces self-oscillation: (**E**) Schematic and (**F**) Overlaid time-lapse snapshots of the beating chain and the extracted skeletons and (**G**) trajectories of individual units. (**H**) Beating frequency f versus speed U of a single unit. (**I**) Limit cycles. (**J-L**) Switching between oscillation (blue) and extension (red) mode is achieved by changing field frequency from 17 kHz to 1.7 MHz, as shown by (**L**) the radius of gyration as a function of time. (**M-O**) Undulation of symmetrically clamped chains pushing loads: (**M**) Schematic, (**N**) light microscopy image with overlaid trajectories (10 kHz) and (**O**) corresponding limit cycle. (**P-R**) Asymmetric attachment induces rotation and tumbling: (**P**) Schematic, (**Q**) brightfield microscopy images with trajectories and (**R**) orientation $\theta_3$ and rotation frequency $\omega$ of the third unit (7 kHz). Color coding (C, D, G, I, N) corresponds to chain units (Fig. 1C). Unless noted otherwise, experiments were conducted at 20 V, 17 kHz. Scale bar: 20 μm.

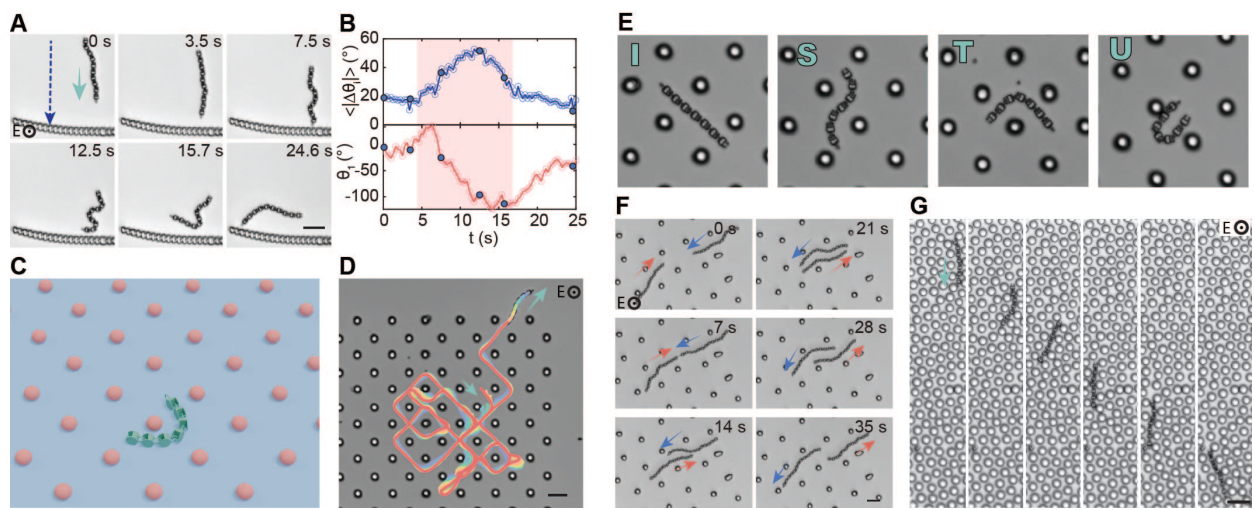

**Fig. 3. Adaptive navigation in complex environments. (A, B)** Upon encountering a wall (red), the chain buckles as measured by the mean local curvature $\langle |\Delta\theta_i| \rangle = \frac{1}{N-1}\sum_1^{N-1} |\theta_{i+1} - \theta_i|$, and reorients the head as captured by the orientation of the head with respect to the tangential direction of the wall $|\theta_1 - \theta_\perp|$ (Movie S7). **(C)** Active, flexible chains adapt their motion to ordered arrays of obstacles (spacing 25 μm) as visible from **(D)** their trajectories, where colors indicate different units of the chain. **(E)** A straight shape is adopted to follow a lane in the pillar array ('I'); s-shapes allow changes of lanes ('S'); bending at 90° and 180° leads to right/left turns (T) and U-turns (U), respectively (Movie S8; 20 V, 20 kHz). **(F)** Their pliable body allows them to move around each other even when confined between obstacles (Movie S9; 20 V, 10 kHz). **(G)** Even in crowded environments of mobile 7 μm silica spheres the chains can easily move by adapting their shape and displacing obstacles as is visible from brightfield microscopy images (Movie S10; electrode separation 250 μm, 20 V, 4 kHz). Scale bar is 20 μm.

# Supplementary Materials for

## Autonomous life-like behavior emerging in active and flexible microstructures

Mengshi Wei, Daniela J. Kraft

Corresponding author: kraft@physics.leidenuniv.nl

**The PDF file includes:**

Materials and Methods
Supplementary Text
Figs. S1 to S6
References (*35–53*)

**Other Supplementary Materials for this manuscript include the following:**

Movies S1 to S10

## Materials and Methods

### 3D microprinting

All structures were designed using AutoDesk Inventor, processed by Describe and fabricated by two-photon polymerization using a commercially available system (Photonic Professional GT, Nanoscribe GmbH). The basic unit of the chains consists of a half-cylinder with diameter 5 µm and thickness 3.5µm. Attached to the curved side of the half cylinder is a handle and to the other side a bar connected to two disk-like extension of the half cylinder's top and bottom (Fig. 1A and Fig. S1A). The flat disks on either side of the bar control and constrain the flexibility of concatenated units to $\pm 50°$ (Fig. 1B).

To attach a load at the front in a symmetric fashion, a hollow standing cylinder (outer diameter 10 µm, inner diameter 6.5 µm, height 11 µm) was connected in a similar handle-bar fashion to the leading unit (Fig. 2, M-O) and Fig. S1B). The load was designed to be light enough for the cylinder to be able to overcome gravity and align its long axis along the field, while at the same time to be large enough to impose drag resistance. For the asymmetrically connected load, the leading unit of a flexible chain was attached off-center to a rectangular prism with side lengths 23.5 µm, 6 µm and 5 µm in a non-flexible manner (Fig. 2P-R and Fig. S1C). Upon activation the longest dimension of the prism aligns with the electric field in the $z$ direction twisting part of the chain that connect to it.

The thus designed concatenated chains were printed on fused silica slides with optimized laser power for printing. To obtain the highest printing resolution, a 63× oil-immersion objective (Zeiss, NA=1.4) together with IP-Dip photoresist (Nanoscribe GmbH) were employed in dip mode. To optimize print fidelity is crucial for replicating the delicate hinges. We printed chains in a one by one consecutive fashion and ensured that each chain was printed within one printing box to avoid drifting of overhanging features. A minimal hatching and slicing distance of 0.1 µm were chosen to obtain a smooth surface.

### Post-printing protocol

Due to the delicate sub-micron thin features of the linkage arms (d=0.5 µm), careful development procedures are essential to avoid deformation by capillary forces that occur during drying. To reduce their effect, we followed a protocol from ref. (*35*): The printed structures were first developed in propylene glycol monomethyl ether acetate (PGMEA, Merck) for approximately 20 minutes, followed by a gentle exchange with isopropyl alcohol (IPA) three times separated by a residence time of 10 min, ensuring that the structures remained submerged to avoid capillary forces. The structures were post-UV cured (Projet CU106) for 90 minutes while remaining submerged in IPA, followed by rinsing with IPA and drying in air. Once dried, the print was sputter-coated with a 50 nm thick silica layer (Coaxial Power System RFG 600) to prevent the units from sticking to each other and to the substrate after release, which is crucial for maintaining the flexibility of the chain.

### Release

Printed chains were removed from the substrate by immersion in MilliQ water and repeated ultrasonication by 10 second pulses until the chains were released. Extensive sonication increases the probability for breaking of the thin links.

## Fabrication of crowded and structured environments

Obstacles and walls (Fig. 3) were printed on ITO-coated glass slides using IP-S resin with a 25× objective (NA 0.8). The wall was designed to have the same width and height (3.5 and 5 $\mu m$) as the chain. The ordered square array of obstacles consisted of 7 × 5 μm oblate ellipsoid arranged with a 25 μm spacing. Oblate ellipsoids were chosen as obstacles because they impose a repulsive force at the equator of the sphere via DEP. The size is chosen such that it is neither too large to distort the electric field significantly nor too small to prevent clamping of the chain due to dipolar attraction in the $z$-direction. Due to the limited printing resolution of the 25× objective, the resulting structure is slightly larger than the design specifications. The size and spacing between obstacles were chosen such that they imposed confinement of the chains, while still allowing U-turns. Curling, caging and hopping of the chains have never been observed (*32–34*). After printing, the structures were developed and treated following the same protocol as was used for the chains.

Crowded environments with mobile obstacles were created by dispersing colloidal silica spheres with 7 μm diameter (microParticles GmbH) inside the sample cell.

## Sample cell design

Experiments were conducted in a cell made by placing a thin ring of silicone (d = 9 mm, height = 120 $\mu m$, Grace, Biolab, SecureSeal) between two ITO-coated slides (surface resistivity 8-12 Ω/sq, Sigma-Aldrich), which were sputter-coated with a 20 nm silica layer prior to use to prevent particle sticking. For experiments, the chains were dispersed in DI water and loaded into the sample cell. To apply an electric field, a small ITO region of each slide was left uncoated for attaching electric wires, which were connected to a function generator (Aim-TTi TG1010A).

## Activation and clamping of the chains

To activate the chains, we apply an electric field with a square wave amplitude with the function generator. The peak-to-peak amplitude and frequency per experiment are indicated in the corresponding captions. Once the electric field is applied, all units align their long axis with the direction of the electric field (z direction) and begin to self-propel on the substrate perpendicular to the field direction (xy plane). At kHz frequencies, the particles move with their curved side leading. Their speed exhibits a linear dependence with $E^2$ and $1/f$ (Fig. S2A). Considering the typical speeds of 1-10 μm/s, the size of a unit ($d = 5$ μm) and the translational diffusion coefficient (D = $4 \times 10^{-14}$ m $\cdot$ s$^{-1}$), we estimate the Peclet number to be on the order of Pe = $vd/D_t$ = 100-1000.

Clamping is achieved by local disturbance of the electric field. Defects on the ITO layer, pollutants on the substrate, or controllably deposited smaller particles can all serve as clamping points. When clamped, rotation of the clamped unit is suppressed fully to a small angular range. The chains can be released from the clamping points by turning off the electrical field.

## Image acquisition and analysis

Samples are imaged using an inverted microscope (Nikon Ti-e) equipped with a 20× long working distance objective (NA=0.75). Videos are captured at 10 frames per second using a sCMOS camera (Teledyne Photometrics Prime, BSI Express) unless stated otherwise.

Image analysis is performed using custom-written Matlab codes. To track the position and orientation of each of the closely linked unit, we threshold the images and calculate the average pixel position to determine their center of mass position in the first frame. The orientation is determined using MATLAB's 'regionprops' function, which calculates the angle of the major axis of an ellipse with the same second moments as the thresholded region. For subsequent frames, we search for features around the positions detected in the previous frame to locate the positions in the current frame. The orientation of each unit not only indicates the local orientation of the chain (as shown in Fig. 1B) but also sets the direction of self-propulsion (or polarization) which is by design inherently along the contour of the chain. Our chains thus classify as polar or tangentially driven active polymers (*26, 27, 30, 31*).

The relative mean squared displacement (MSD) of the units with respect to the leading unit from Fig. 2D is calculated following ref. (*30*):

$$\Delta r_{1,i}^2(t, \tau) = \langle [r_1(\tau) - r_i(\tau + t)]^2 \rangle,$$

where $r_1$ is the position of the first, or leading, unit in the chain and $r_i$ is the position of the $i^{th}$ unit counted from the leading unit until the last, $N^{th}$, unit, $\tau$ is time and $t$ the time difference between which the positions of the first and $i^{th}$ unit are compared, where $\langle ... \rangle$ denotes an average over all frames.

The orientation of the units of self-oscillating chains is calculated as the orientation with respect to a reference orientation. In the clamped case, the orientation of the clamped unit is taken as reference. During load pushing, the reference orientation is defined by the tangential direction of the load's trajectory after smoothening it using a Savitzky-Golay filter.

## Supplementary Text

### Bending stiffness

We extract the angular spring constant of a Brownian two-unit hinge by fitting the probability density function of the relative angle with a Gaussian function (Fig. 2B). From the width $\sigma$ of the fitted Gaussian, we calculate the angular spring of the passive chain as $k = \frac{k_B T}{\sigma}$. This stiffness has a purely entropic origin caused by the geometrically constrained range of motion, similar to freely jointed colloidal structures (*36*).

For the active chains, we extract the persistence length using the software package "easyworm".

### AC field induced propulsion mechanism

Colloidal particles in alternating current (AC) electric field can propel due to a range of electrokinetic mechanisms, include electrohydrodynamics (EHD) (*25, 37*), induced-charge electrophoresis (ICEP) (*17, 38, 39*) and self-dielectrophoresis (sDEP) (*17, 40, 41*). All three mechanisms may contribute to the particle motion, with the strength of their contributions depending on various factors such as the field strength, frequency, salt concentration, particle design and material (*40–42*). For metal-dielectric Janus particles (JPs) with a size similar to the unit size of the chains employed here, i.e. from $5 - 10 \, \mu m$, suspended in low-conductivity solutions ($10^{-5} - 10^{-3}$ M KCl) EHD typically occurs at 10–100 Hz, ICEP dominates around 1–10 kHz, and a transition from ICEP to sDEP is seen at approximately 10–100 kHz.

We measure the frequency dependent dynamics of a single unit and find high velocities in the curved-side leading direction occurring between about 1 kHz and 150 kHz, see Fig. S2B. Below and above this frequency range, the units propel in the opposite direction with their flat side leading and with significantly lower velocities. Another direction reversal towards propulsion with curved side leading was found at 800 kHz, again at very low velocities as shown in Fig. S2B. At the same time, the particle velocity scales with the electric field strength as $U \propto E^2$ independent of the chain length, see Fig. S2A.

The frequency sensitivity of EHD and ICEP stems from the charge relaxation time of the electric double layer induced on the electrolyte-electrode and electrolyte-particle surfaces, respectively, and only differs by the distance relevant for the charging, i.e. the distance between the electrodes $2H = 120$ μm and the size of the self-propelling particle $a = 5$ μm itself, respectively: $f_{\text{EHD}} = \frac{D}{2\pi\lambda_0 H}$ and $f_{\text{ICEP}} = \frac{D}{2\pi\lambda_0 a}$, where $D$ is the ionic diffusivity of the electrolyte and $\lambda_0$ denotes the Debye length (43). For typical experimental conditions presented here, the ionic diffusivity can be estimated as $D \approx 5.3 \times 10^{-9}$ m$^2 \cdot$ s$^{-1}$, accounting for the dissolved atmospheric $CO_2$, which forms $HCO_3^-$ and $CO_3^{2-}$ in Milli-Q water at pH 5.6 (40). The Debye length can be estimated as $\lambda_0 = \sqrt{\frac{\epsilon_m RT}{2F^2 z^2 c_0}} \approx 194$ nm, where $\epsilon_m = 80$ is the permittivity of water, $R$ is the gas constant, $T$ is the absolute temperature, $F$ is the Faraday constant, $z$ is the ionic valence, and $c_0 = 2.5 \times 10^{-6}$ M is the concentration of ions estimated from the pH. For our system, the frequencies for EHD and ICEP can then be estimated as $f_{\text{EHD}} \approx 73$ Hz and $f_{\text{ICEP}} \approx 1.7$ kHz.

The generation of EHD flow moreover requires the presence of a nearby electrode where the presence of the particle distorts the local electric field, thereby inducing a flow of the electric double layer which propels the particle. To determine whether particle activity in our system depends on proximity to the substrate, we levitated the chains into the bulk fluid, away from any bounding surfaces, using optical tweezers (Fig. S3). Upon switching the optical tweezers off, the chains settled due to gravity. Once an AC electric field (20V, 10 kHz) was applied, the chains aligned again and exhibited directed motion with their curved side leading (Fig. S3, A and B). Remarkably, there was no significant difference in the propulsion speed in bulk compared to the speed observed near the electrode (Fig. S3C). This observation provides strong evidence that EHD flow is not necessary for self-propulsion at kHz frequencies.

Unlike EHD, ICEP generates electroosmotic flow directly around the particle. Differences in polarizability such as is the case for dielectric particles half-coated with metal (39) but also an anisotropic particle shape alone (44-46) can induce differences in the flow velocity, resulting in net particle motion. Ideally polarized, nearly symmetric particles, have been predicted to move with their pointed end forward (45), and this behavior has been experimentally observed in quartz particles (47, 48). Although dielectric polymer-based materials like the one used here have significantly lower electro-osmotic flow speeds compared to metal, these observations align with our experimental results, where half-cylinder particles consistently move with their pointed side leading.

In addition to ICEP, self-dielectrophoresis (sDEP) (might contribute to particle propulsion (22). sDEP refers to the movement of a particle within a nonuniform electric field generated by the particle itself which arises from the difference in dielectric properties between the particle and the surrounding medium. The induced dipole on the particle experiences a dielectric force due to the self-induced field gradient, causing the particle to move either towards the high electric field region (positive dielectrophoresis) or the low electric field region (negative dielectrophoresis). Positive sDEP has been shown to drive the movement of Janus particles with their metal cap leading at MHz frequencies (17, 40, 49), and metal sheets (50). Furthermore, dielectrophoresis has been utilized to assemble particles into colloidal molecules (25) and micro-cars (24). To identify whether our particles undergo negative or positive DEP, we calculate the dependence of the Clausius-Mossotti factor $K$ on the field frequency as (51, 52): $K = \frac{\epsilon_p^* - \epsilon_m^*}{\epsilon_p^* + 2\epsilon_m^*}$, where $\epsilon_{p,m}^* = \epsilon_0 \epsilon_{p,m} - i\frac{\sigma_{p,m}}{\omega}$ and $\sigma_{m,p}$ are the electric conductivities of the medium and the particle, respectively. For our system, $\epsilon_m = 80$, $\epsilon_p = 4$, $\sigma_m = 4 \times 10^{-5}$ S $\cdot$ m$^{-1}$, and $\sigma_p = $ $10^{-14}\,\mathrm{S} \cdot \mathrm{m}^{-1}$. Particle properties were chosen to be those of polymethylmethacrylate which is a major component of the employed photoresist. The negative value of $\mathrm{Re}[K]$ for all frequencies indicates negative dielectrophoresis, since the force is given by $F_{DEP} = 2\pi a^3 \epsilon_0 \epsilon_m \, \mathrm{Re}[K] \Delta |E|^2$ (Fig. S4A).

To simulate the dielectrophoresis (DEP) of our particles driven by self-generated electric field gradients, we utilized COMSOL Multiphysics. The simulations were performed using a 2D model incorporating the Electrostatics (es) module under DC field conditions. In this model, a half-disk particle (radius $a = 2.5\ \mu m$, permittivity $\epsilon_p = 4$, conductivity $\sigma_p = 10^{-14}\,\mathrm{S} \cdot \mathrm{m}^{-1}$) was placed within a square water domain (120 $\mu$m $\times$ 120 $\mu$m, permittivity $\epsilon_m = 80$, conductivity $\sigma_m = 4 \times 10^{-5}\,\mathrm{S} \cdot \mathrm{m}^{-1}$). An electrical potential of 20 V was applied across two opposing boundaries of this domain. The resulting electric field distributions, both for particles in bulk solution and those near the substrate, indicates higher field strengths at the particle's equator and lower field strengths above and below the equator along its curved surfaces (as shown in Fig. S4B). Consistent with negative dielectrophoresis (nDEP), our particles are expected to migrate towards regions of lower field strength. To quantify this, we calculated the net DEP force along the x-direction ($\mathrm{F}_x$). This was achieved by integrating the Maxwell stress tensor within the near-field approximation, following the approach detailed by (*23*). The calculated integrated forces were found to be $7.12 \times 10^{-7}\,\mathrm{N} \cdot \mathrm{m}^{-1}$ and $4.26 \times 10^{-7}\,\mathrm{N} \cdot \mathrm{m}^{-1}$ for particles in bulk and near the substrate, respectively, the unit $\mathrm{N} \cdot \mathrm{m}^{-1}$ is due to the 2D nature of the simulation. The positive value of the forces corresponds to the movement in the x-direction and thus with the curved side leading. We estimated the corresponding particle speed (U) by balancing the calculated net force ($\mathrm{F}_x$) with the Stokes drag force. For a half cylinder with a depth d = 3.5 $\mu$m and an effective radius r = 2.2 $\mu$m, the drag force is given by $\mathrm{F}_x = 6\pi \eta r U$, where $\eta$ is the fluid viscosity. This calculation yields speeds of 6.8 $\mu$m/s and 4.0 $\mu$m/s for particles in bulk and near the substrate, respectively, which closely align with our experimental results.

A key difference between ICEP and DEP is that ICEP generates an electro-convective flow, while DEP involves the directed transport of electrically polarizable particles. To further distinguish between potential driving mechanisms, we introduced tracer particles (600 nm fluorescent silica particles) to investigate the presence of a flow field around our active unit (Fig. S4C). A single active unit (green fluorescent) was aligned with the electric field and immobilized on the substrate, allowing us to examine interactions in the plane parallel to the substrate. Prior to applying the electric field, the tracer particles were homogeneously distributed within the field of view and exhibited Brownian motion. Upon turning on the electric field (at a typical operating frequency of 10 kHz), the tracer particles were attracted to the cavity of the half-cylinder of the curved side instantaneously and remained trapped, without being ejected. This is because the used tracer particles experience negative dielectrophoresis and thus were attracted to the low field strength region, as shown in the simulation (Fig. S4B). When the electric field was turned off, the trapped tracer particles were released from the active unit and dispersed back into the bulk solution. This observation suggests the absence of significant fluid flow and confirms the dominance of DEP forces induced by the active particle. We thus conclude that the velocity peak at 1-100 kHz contains contributions from both ICEP and sDEP, but with dominant contributions stemming from sDEP. The mechanism for the propulsion inversion at even higher frequencies (1 MHz) requires further research for elucidation.

## Dipolar interactions

The electric field $\vec{E}$ induces the formation of dipoles $\vec{p} = 3V_p\epsilon_0\epsilon_m\text{Re}[K]\vec{E}$, where $V_p$ is the volume of the particle and $\epsilon_0$ the vacuum permittivity. This leads to a dipolar interaction between induced dipoles, characterized by a dipole interaction energy

$$U = \frac{p_1 p_2 (1 - 3\cos\theta)}{4\pi\epsilon_0\epsilon_m r^3},$$

where $p_{1,2}$ are the induced dipoles, $\theta$ is the angle between the center-center vector $\vec{r}$ and the electric field $\vec{E}$. For parallel dipoles the interaction energy is found to be repulsive for all frequencies, in line with the straightening and expansion of the chains upon turning the electric field on, see Fig. S4A (bottom).

## Self-oscillations and bending rigidity

Upon clamping, the chain buckles due to the active force and undergoes self-oscillations. The frequency of these oscillations just like the precise path followed during the oscillations depends on the applied voltage as shown in Fig. 2 and 8A-F. We here expand on these observations and their origin.

The linear scaling of the oscillation frequency $f \propto U$ with the propulsion speed of a single unit $U$ observed in our experiments differs from previous theoretical predictions for tangentially driven active polymers, which suggested an exponent of 3/4 (26). This result was derived by modeling the polymer as an elastic filament with a constant elastic bending rigidity $\kappa$ and assuming that the active energy, which is first transferred into elastic bending energy, is subsequently dissipated by viscosity. A different result was obtained in experiments on ICEP-driven Janus particles assembled into beating flagella, where a linear scaling of the beating frequency with the propulsion speed was found (18). The deviation from the theoretical prediction was rationalized by a variation of the bending rigidity with the activation voltage.

In our system, the geometric constraints imposed by the concatenation, induced dipoles and thermal nature of our chains all introduce bending rigidities. Firstly, the design imposes a sterically determined maximum bending angle $\Delta\theta_{max}$ between two adjacent units. Above $\Delta\theta \geq \Delta\theta_{max}$, this geometric limit imposes an infinite bending resistance $\kappa(\Delta\theta \geq \Delta\theta_{max}) = \infty$. For angles below the maximum, an effective bending rigidity is introduced in two additional ways: In the passive state, the geometric constraint induces an effective bending rigidity $\kappa_p(\Delta\theta < \Delta\theta_{max})$ which originates from thermal fluctuations and can be extracted from the probability distribution of the angle between two connected units, $p(\Delta\theta)$, shown in Fig. 1(B) (36). In thermal equilibrium, the probability distribution follows a Boltzmann distribution and can be fitted to a Gaussian distribution:

$$p(\Delta\theta) = \sqrt{\frac{\kappa_p}{2\pi k_B T}} \exp\left(-\frac{\kappa_p}{2k_B T}\Delta\theta^2\right)$$

Thus, $\kappa_p = \frac{k_B T}{\langle\Delta\theta^2\rangle}$, which is estimated to be $2.3 \times 10^{-20}\,\text{N}\cdot\text{m}^2$. From the probability distribution, we can furthermore extract the maximum bending angle $\Delta\theta_{max}$ for this specific geometry to be about 50°.

Secondly, in the active state, the chain expands due to repulsive dipolar interactions between the units, see Sec.4 and Fig. S4A. Particles with the same material and thus the same $Re[K]$ which are aligned side by side (i.e. $\vec{r}$ is perpendicular to $E$ and $\theta = \pi/2$) always experience a repulsive force, and the interaction scales with $E^2$. This dipolar repulsive force between adjacent units effectively increases the chain's rigidity by promoting stretching. The bending rigidity induced by dipole-dipole interactions scales as:

$$\kappa(\Delta\theta < \Delta\theta_{max}) = \kappa_{\text{d-d}} \sim \frac{E^2}{r^3}$$

During the self-oscillations, the units at different positions relative to the clamping point experience different degrees of bending. The unit closest to the clamping point experiences the largest compression force, therefore bending the most and reaching the extreme angle $\Delta\theta_{\max}$. Units in the rear part may not experience compression and instead exhibit a railway-like motion without bending.

The conformation and maximum lateral extent during buckling and self-oscillation, however, remain similar across different activation field strengths and thus active forces (Fig. S5, A-F). The reason for this is that the active forces (even the one driven by the lowest voltage at our experiments), are always much larger than the resistance to buckling, leading to the buckled conformation being determined by the maximally possible bending angle and not by a competition between the active force and the thermal or dipolar interaction induced bending rigidities. Therefore, neither the scaling derived from an active polymer description, nor the scaling found in assemblies of ICEP-driven Janus particles describes the scaling of the oscillation frequency with the speed found in our case. Instead, the maximum relative bending angle ($\Delta\theta_{\max}$) set by the connections dictates the buckling profile and the oscillation frequency thus scales simply with the velocity of the single unit.

Still, we can observe an effect of the dipolar repulsion on the conformation of the chain: at low activation voltage, dipolar interactions are weak, and thermal fluctuations still affect the motion of the individual units. With increasing field strength, the chain shows a slight expansion indicating an increase in bending rigidity which results from dipolar repulsion between the units. Indeed, the variance in the angle between two units decreases as the activation voltage increases (Fig. S5H), indicating that the effective bending rigidity resulting from dipole interactions increases. At higher voltages, the buckling decreases again, likely due to a larger contribution from the stronger active force.

Similarly, the persistence length of the chain shows an initial increase in persistence length with increasing voltage, followed by a decrease (Fig. S5I). We estimated the persistence length by measuring the mean square end-to-end distance using the method derived from the worm-like chain (WLC) model for semi-flexible polymers (*53*). These measurements suggest, albeit minor, a decrease in buckling, followed by a subsequent increase with increasing field strength, reflecting again the interplay between active force and buckling.

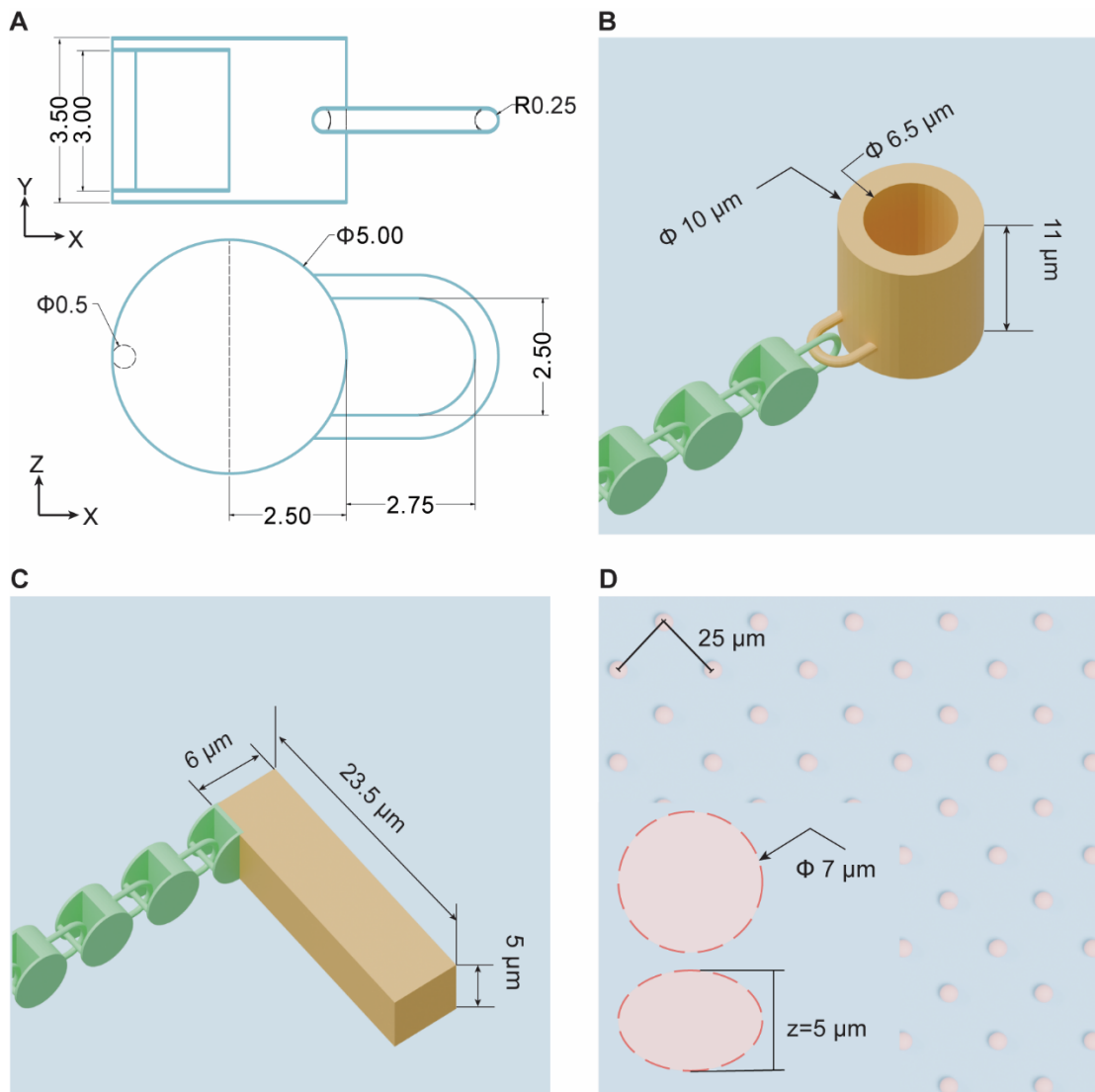

**Fig. S1. Auto CAD design of 3D printed structures.** (**A**) Single unit design, with all length scales in μm. (**B**) Design of chain with a symmetric load: a chain is flexibly connected to a hollow cylinder without no rotational constrain imposed from the connection. (**C**) Design of chain with an asymmetric load: the head of the chain is fixed to one end of a rectangular prism. (**D**) Design of the square array of obstacles with a spacing of 25 μm. Each obstacle is an oblate ellipsoid with 7 μm major and 5 μm minor axis.

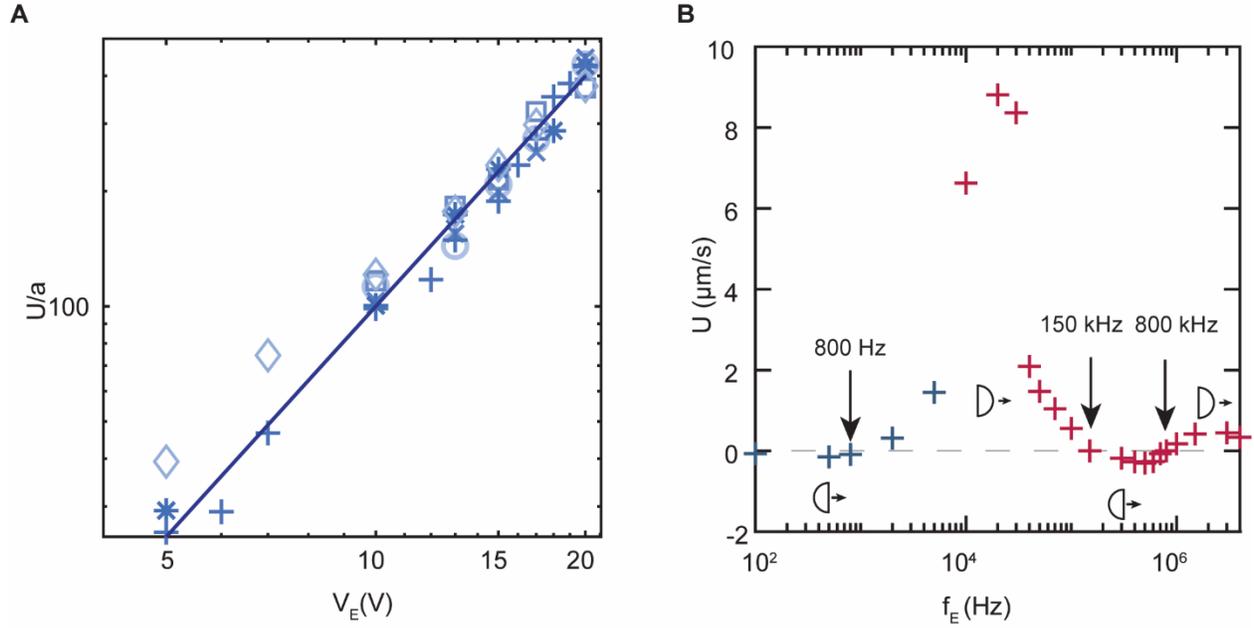

**Fig. S2. Dependence of the particle swimming velocity on electric field frequency and strength.** (**A**) The speed of the chain **U**, normalized by a prefactor **a**, follows a trend consistent with $\mathbf{V_E^2}$, clearly demonstrating a linear relationship between the speed and the square of the field strength, regardless of the variations in total number of segments $\mathbf{N_s}$ and field frequency $\mathbf{f_E}$ explored. Different symbols refers to different experiments with the following conditions: ○: $\mathbf{N_s =}$ **1**, $\mathbf{f_E =}$ **10 kHz**, **a** = 0.0146, ◊: $\mathbf{N_s =}$ **1**, $\mathbf{f_E =}$ **17 kHz**, **a** = 0.0182; ▫: $\mathbf{N_s =}$ **8**, $\mathbf{f_E =}$ **10 kHz**, **a** = 0.0435; ×: $\mathbf{N_s =}$ **8**, $\mathbf{f_E =}$ **20 kHz**, **a** = 0.0431; +: $\mathbf{N_s =}$ **11**, $\mathbf{f_E =}$ **17 kHz**, **a** = 0.0435; ✳: $\mathbf{N_s =}$ **13**, $\mathbf{f_E}$ =10 kHz, **a** = 0.0354, solid line indicates scaling with $V_E^2$. (**B**) Dependence of the velocity **U** of a single unit on field frequency. Experiments in this study were conducted at 1-150 kHz, where the half cylinder moves with its curved face leading (positive **U**). At lower and higher frequencies, the chain moves with its flat side leading (negative **U**). The direction of motion reverses again from negative to positive at 800 kHz. These reversals in direction reflect a complex interplay of various electro-kinetic phenomena. Blue symbols at low frequency were obtained at $\mathbf{V_E}$ = 10 V to avoid hydrolysis of the aqueous solvent, red symbols were obtained at $\mathbf{V_E}$ = 20 V.

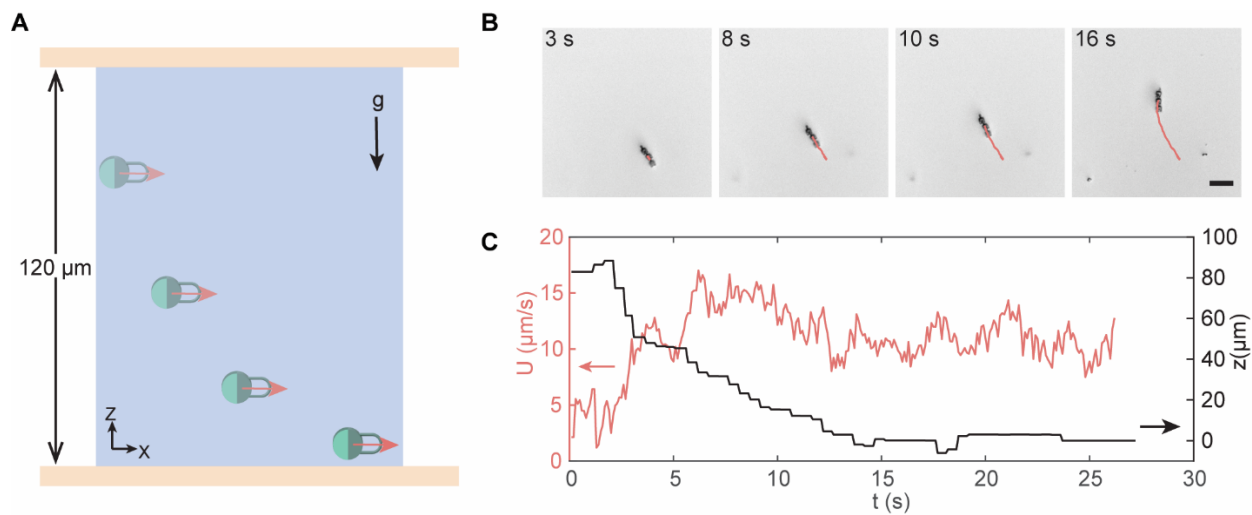

**Fig. S3. Propulsion in bulk**. (**A**) Schematic illustrating an active unit undergoing forward motion in bulk during sedimentation after having been lifted away from the substrate using optical tweezers. (**B**) Bright field microscopy snapshots, overlaid with the trajectory of a 3-unit chain moving in bulk during sedimentation (20 V, 10 kHz). Scale bar: 20 μm. (**C**) Instantaneous speed of the chain (black curve) and its distance to the bottom substrate (orange curve).

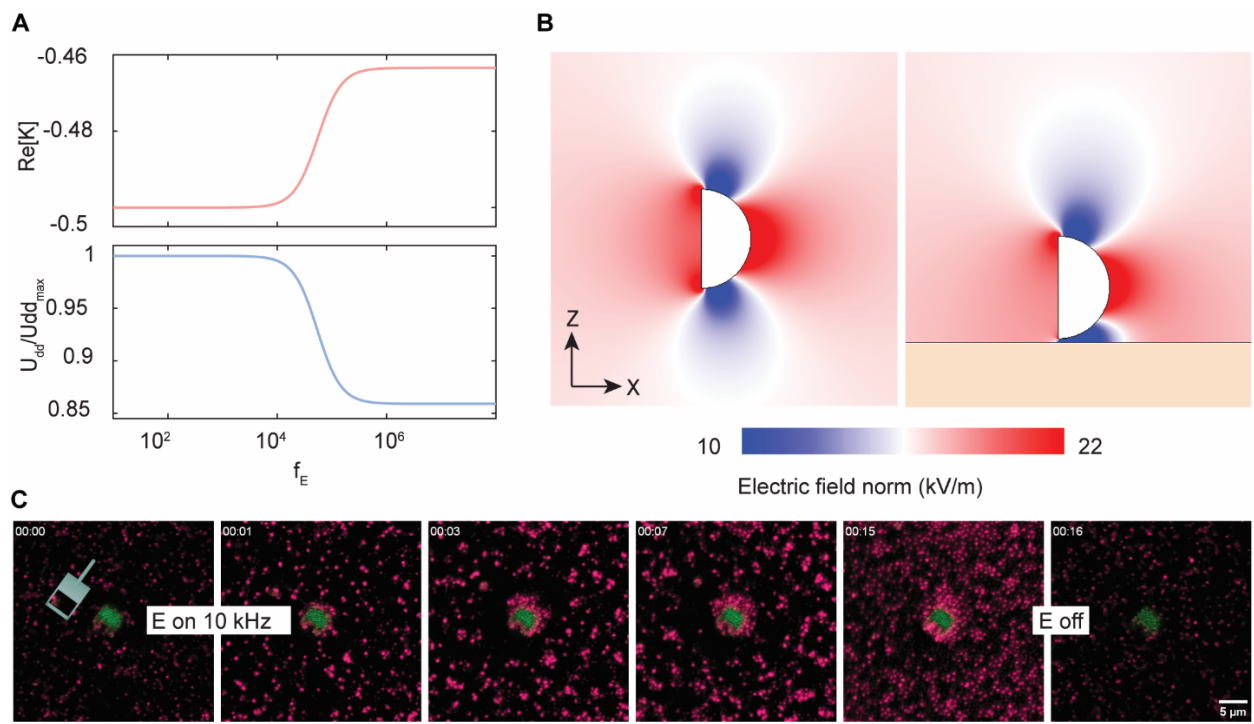

**Fig. S4. Evidence for sDEP motion.** (**A**) Dielectric spectra and interactions: the real part of the complex Clausius-Mossotti factor (top) and the dipole-dipole interaction normalized by the maximum value within the frequency range investigated (bottom). (**B**) Non-uniform electric field generated around a semicircle ($\epsilon_p = 4$) placed in DI water ($\epsilon_m = 80$) in bulk (left panel) and near the substrate (right panel, 400 nm above the substrate), as simulated by COMSOL (20 V, 120 **μm** electrode distance). (**C**) Time-lapse snapshots illustrating the dynamic response of 600 nm silica tracer particles near an immobilized active unit. When the electric field is turned on (20 V, 10 kHz), tracer particles (magenta) are attracted to and captured around the active unit (green). Upon turning off the electric field, the captured tracers are released and re-dispersed into the bulk solution.

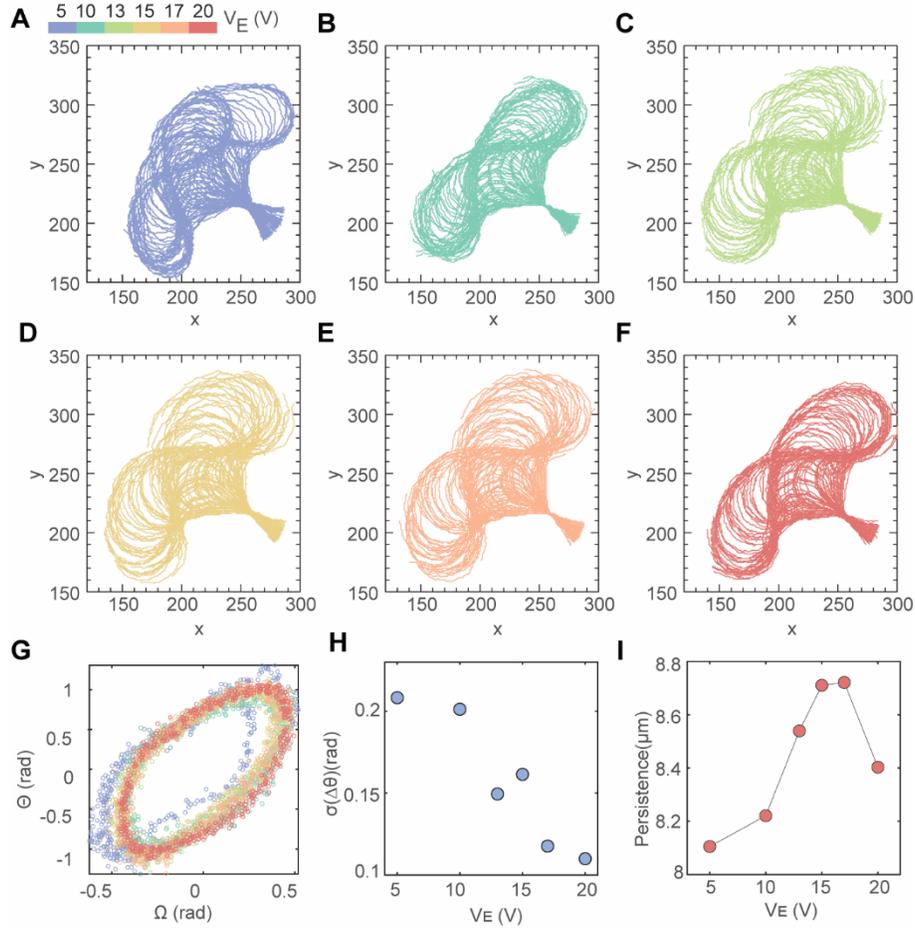

**Fig. S5. Self-oscillation under varying field strength.** (**A-F**) Skeletons of the clamped chain for different electric voltages, corresponding to the data in Fig. 2, show a similar conformation, indicating that it is imposed by geometric constrains stemming from the design of the unit. The colors correspond to different electric field voltages as indicated in the color bar. (**G**) Mean curvature $\boldsymbol{\Theta}$ versus mean polarization $\boldsymbol{\Omega}$ of the chain for different voltages (using the same color scheme as in (A)). Similar limit cycles are executed for all applied voltage, except for $\boldsymbol{V_E}$ = 5 V, where the chain is floppier, resulting in greater fluctuations. (**H**) Standard deviation of angle fluctuations of a 2-unit chain for different voltages, indicating an increase in bending rigidity with increasing voltage. (**I**) Persistence length of the chain extracted from the skeletons in (A) and computed following Ref. (58), indicating the competing effects of dipole repulsion and active compression, both of which increase with field strength.

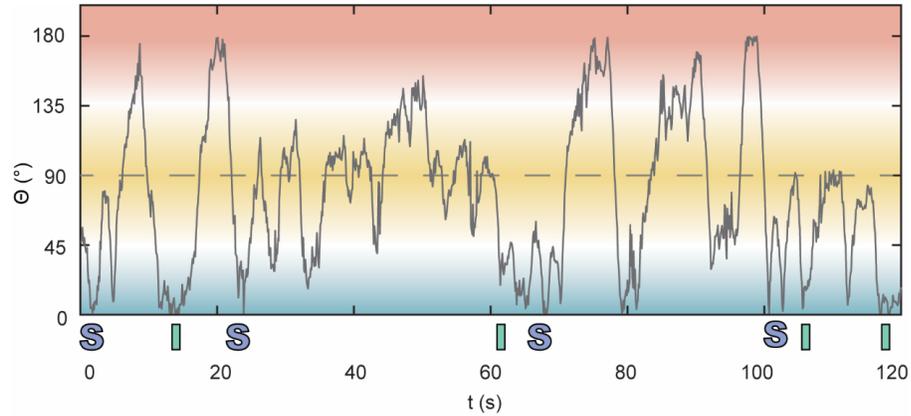

**Fig. S6 Time evolution of the chain's curvature during navigation through the obstacle array.** Active chains navigate squared obstacle arrays by adopting four specific conformations corresponding to different values of the chains curvature $\Theta$ : $\Theta = 0$ (light blue) can correspond to either straight motion ('I') or s-shapes which allow switching between lanes ('S'); $\Theta = 90°$ corresponds to right and left turns (yellow), $\Theta = 180°$ corresponds to U-turns (red). The conformation of the chain constantly switches between different conformations as illustrated by the time evolution of the mean curvature.